\documentclass[letterpaper,12pt]{article} 
\usepackage{amsmath,amsfonts}
\usepackage{graphicx}
\usepackage{latexsym}

\providecommand{\e}[1]{\mspace{2mu}\mathrm{e}^{#1}\mspace{1mu}}
\providecommand{\sgn}[1]{\mspace{2mu}\mathrm{sgn}\mspace{2mu}#1\,}
\providecommand{\I}{\mathrm{I}}
\providecommand{\K}{\mathrm{K}}
\providecommand{\J}{\mathrm{J}}
\providecommand{\N}{\mathrm{N}}
\providecommand{\mpl}[1]{M_{Pl(#1)}}

\title{Geodesics and Newton's Law in Brane Backgrounds}
\author{W.~M\"uck\thanks{E-mail address: \texttt{wmueck@sfu.ca}},
K.~S.~Viswanathan\thanks{E-mail address: \texttt{kvisawana@sfu.ca}}
and I.~V.~Volovich\thanks{Permanent address: \emph{Steklov Mathematical
Institute, Gubkin St.~8, 117966 Moscow, Russia},
\texttt{volovich@mi.ras.ru}}\\  
\small \emph{Department of Physics, Simon Fraser University, Burnaby,
B.C., V5A 1S6 Canada}}

\begin{document}
\maketitle
\begin{abstract}
In brane world models our universe is considered as a brane
imbedded into a higher dimensional space. We discuss the behaviour of
geodesics in the Randall-Sundrum background and point out that free
massive particles cannot move along the brane only. The brane is
repulsive, and matter will be expelled from the brane into the extra
dimension. This is rather undesirable, and hence we study an
alternative model with a non-compact extra dimension, but with an
attractive brane embedded into the higher dimensional space.  
We study the linearized gravity equations and show that Newton's
gravitational law is valid on the brane also in the
alternative background. 
\end{abstract}
\newpage

\section{Introduction}
It has recently been suggested  by Randall and Sundrum 
\cite{Randall99b} that four-dimensional gravity can arise at long distances on
a brane embedded in a five-dimensional anti-de Sitter space. In their
model the fifth dimension is non-compact. An effective dimensional reduction
occurs because the metric perturbations admit a bound state
solution which looks like a four-dimensional graviton bound to the brane.
Earlier work appeared in \cite{Rubakov83,Visser85,Akama00}.
\nocite{Arkani00,Goldberger99,Gogberashvili99,Behrndt99,Halyo99,Chamblin99,deWolfe99,Arefeva99}
\nocite{Csaki99,Charmousis99,Garriga99,Cohen99,Verlinde99,Lykken99,Kaloper99}
\nocite{Sasaki99,Ivanov99,Myung00,Youm99,Youm00a,Youm00b,Kallosh00}
\nocite{Csaki00,Giddings00,Gregory00}
This interesting alternative to compactification has been discussed
in a number of recent papers \cite{Arkani00}--\cite{Gregory00}.

The metric of the Randall-Sundrum (RS) background has the form
\begin{equation} \label{metric1}
 ds^2 = e^{-2k|y|} \eta_{\mu\nu} dx^\mu dx^\nu + dy^2, 
\end{equation}
where $\eta_{\mu\nu}=\text{diag}(-1,1,1,1)$ ($\mu,\nu=0,1,2,3$).
It was argued in \cite{Randall99b} that the Kaluza-Klein  excitations,
although they are light, are suppressed near the brane and almost
decouple from the matter fields. Moreover, it is assumed that 
matter fields are trapped to the brane by a certain mechanism.

If the brane describes our Minkowski space-time, then there should
exist trajectories for free massive particles located on the brane
only. However, this is not true for the metric
(\ref{metric1}).  
The trajectory of the free particle in the metric (\ref{metric1})
has the form (see Sec.~\ref{geodesics} for details, here we take
$y_0=\dot y_0=0$) 
\begin{equation} \label{traj}
  x^{\mu}=x_0^{\mu}+v^{\mu}t, \quad y=\frac{1}{2k}\ln(1-v^2k^2t^2)
\end{equation}
Therefore, free massive ($v^2<0$) particles cannot move in Minkowski
space-time without being ineviatably expelled
into the $y$-dimension. 
This seems rather undesirable, even if for a very small 
(Planck scale) $k$ the time needed for a significant deviation in
the fifth direction will be rather large. Thus, in the RS background,
some other, non-gravitational mechanism is needed in order to trap
matter on the brane. The simplest way to obtain an attractive
brane would be to change the sign of the brane tension, although one
might argue that this would imply other 
undesirable features. This alternative was considered in
\cite{Visser85,Gogberashvili98a,Gogberashvili98b,Bajc99}.

In the present paper, we shall study the RS background and
the alternative possibility, whose background metrics are given by 
\begin{equation} \label{metric}
 ds^2 = \e{\mp2k|y|} \eta_{\mu\nu} dx^\mu dx^\nu + dy^2. 
\end{equation}
The upper or lower signs correspond to the RS and alternative
backgrounds, respectively. This convention shall be used throughout this
paper, and $k$ is positive, $k>0$.
The metric \eqref{metric} is a solution to Einstein's equation for the
action 
\begin{equation}
\label{action}
  S = \int d^4 x \int\limits dy\, \sqrt{-g} (R-2\Lambda) +
  \sigma \int_{y=0} d^4 x \sqrt{-g_B},
\end{equation}
where the cosmological constant and brane tension are 
\begin{equation}
\label{lambda}
  \Lambda = -6k^2, \qquad \sigma = \mp 12k.
\end{equation}

We shall now give a brief outline of the rest of the paper. 
First, in Sec.~\ref{geodesics} we study the geodesics in the two
backgrounds and find that only in the alternative background gravity 
provides a mechanism for the trapping of matter on the brane. 
Secondly, in Sec.~\ref{lingrav} we consider the linearized
gravity equations, which shall be used in Sec.~\ref{newton} to derive
the Newtonian limit for gravity on the brane. 
We find that in both backgrounds the
gravitational potential for a static point source will be $\sim-1/r$,
and we find an exact formula for the corrections. In Sec.~\ref{modes},
we give expressions for the graviton modes in the both
backgrounds. Space-like modes are absent in the RS background, but our
results are inconclusive for the alternative background.

\section{Geodesics}\label{geodesics}
In this section, we explicitely solve the geodesic equation in the RS
and alternative backgrounds. We shall find that, in the RS background,
ordinary matter will be expelled from the brane, but in the
alternative background, the brane is attractive.

Using the zeroth order terms of the connections given in the appendix,
the geodesic equation takes the form
\begin{align} 
\label{geod1}
 \frac{d^2 x^\mu}{d\theta^2} \mp 2k \sgn{y} \frac{dx^\mu}{d\theta}
 \frac{dy}{d\theta} &=0,\\
\label{geod2}
 \frac{d^2 y}{d\theta^2} \pm k \sgn{y} \e{\mp2k|y|} 
 \eta_{\mu\nu} \frac{dx^\mu}{d\theta}
 \frac{dx^\nu}{d\theta} &=0.
\end{align}

We start by integrating eqn.\ \eqref{geod1}, which yields
\begin{equation}
\label{geod1sol}
  \frac{d x^\mu}{d\theta} = v^\mu \e{\pm 2k|y|},
\end{equation}
where $v^\mu$ is a constant four-vector. Eqn.\ \eqref{geod2} is
explicitely solved by the first integral of the geodesic equation,
 \[ \left( \frac{dy}{d\theta} \right)^2 + \e{\mp2k|y|} 
 \eta_{\mu\nu} \frac{dx^\mu}{d\theta}
 \frac{dx^\nu}{d\theta} = C, \]
where $C$ is a constant. Hence, inserting eqn.\ \eqref{geod1sol}, we
find 
\begin{equation}
\label{geod2sol}
  \left( \frac{dy}{d\theta} \right)^2 + \e{\pm2k|y|} v^2 = C,
\end{equation}
where $v^2=v^\mu v^\nu \eta_{\mu\nu}$. 

It is convenient to change the parameterization to a non-affine
parameter $t$ such that
\[ \frac{dt}{d\theta} = \e{\pm2k|y|}.\]
Then, eqn.\ \eqref{geod1sol} becomes (notation $\dot x = dx/dt$)
\begin{equation}
\label{xsol}
  \dot x^\mu = v^\mu \qquad \Rightarrow \qquad x^\mu = x_0^\mu +v^\mu t,
\end{equation}
which shows that we can choose a reference frame such that $t=x^0$,
i.e.\ $t$ is the time on the brane.
Moreover, eqn.\ \eqref{geod2sol} becomes
\begin{equation}
\label{yeq}
  \dot y^2 \e{\pm4k|y|} + \e{\pm2k|y|} v^2 = C, 
\end{equation}
and we can determine $C$ from the initial data:
\begin{equation}
\label{C} 
 C= \dot y^2_0 \e{\pm4k|y_0|} + v^2 \e{\pm2k|y_0|}.
\end{equation}
Before integrating eqn.\ \eqref{yeq}, we note that it depends only on
$|y|$, as long as we do not pass through $y=0$. Therefore, it is
sufficient to consider $y>0$; replacing $y$ with $|y|$ at the end will
take care of the case $y<0$. For $y>0$, we can integrate eqn.\
\eqref{yeq} and find
 \[ \sqrt{C - v^2 \e{\pm 2ky}} = \pm v^2 k t + \sqrt{C-v^2 \e{\pm
  2ky_0}}, \] 
where we have again expressed the integration constant by the initial
data. Notice, that the $\pm$ sign in front of the term containing $t$
on the right hand side is not (yet) related to the sign in the
exponentials, but stems from the ambiguity of taking a square
root. After some simple steps involving the substitution of $C$ from
eqn.\ \eqref{C} we obtain
 \[ \e{\pm 2ky_0} - \e{\pm 2ky} = v^2 k^2 t^2 \pm |\dot y_0| 2k
 \e{\pm 2ky_0} t. \]
From the initial data we can deduce that we have to
replace $\pm |\dot y_0|$ by $\mp \dot y_0$. Thus, the final result is 
\begin{equation}
\label{y}
 \e{\pm2k|y|} = \e{\pm2k|y_0|} \pm 2k\dot y_0 
 \e{\pm2k|y_0|} t - v^2 k^2 t^2.
\end{equation}
The solution \eqref{y} is valid, as long as $|y|\ne 0$.

If we hit the brane at $y=0$, we have to match a solution for $y>0$
with a solution for $y<0$. However, from eqn.\ \eqref{geod2} we see
that the velocity $\dot y$ must be continuous at $y=0$, since the
second term in that equation is finite. Thus, the brane will not
deflect particles gravitationally, but we might expect that
non-gravitational interactions with matter on the brane might do so. 

\begin{figure}[ht]
\begin{center}
\includegraphics[width=.5\textwidth]{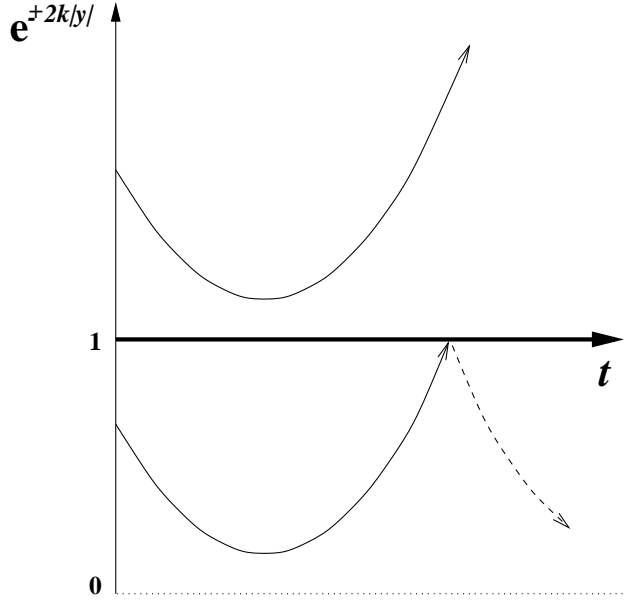}
\end{center}
\caption{\label{geodfig}
\small The solution \eqref{y} for ordinary matter, $v^2<0$. 
The region $\e{\pm 2k|y|}>1$
corresponds to the RS background, the region
$\e{\pm 2k|y|}<1$ to the alternative background. In both cases, the
brane sits at $\e{\pm 2k|y|}=1$.} 
\end{figure}

The interpretation of the solution \eqref{y} is rather simple: In the
RS background, which corresponds to the upper sign, ordinary
matter ($v^2<0$) is expelled from the brane. On the other hand, tachyonic
particles ($v^2>0$) are attracted to the brane, whereas massless
particles are not affected by its presence. Using the lower sign, the
brane attracts ordinary matter. This is sketched in Fig.~\ref{geodfig}.

Another interesting fact is that, for the alternative background, 
$\e{-2k|y|}=0$ corresponds to $|y|=\infty$. Thus, tachyonic particles will be
expelled to $|y|=\infty$ in finite brane time, as will massless particles
with the right initial conditions. Moreover, there exist initial
conditions for ordinary particles, which will yield $|y|=\infty$ in
finite brane time. 

\section{Linearized Gravity} \label{lingrav}
In this section, we shall study the linearized gravity equations with
two applications in mind: The derivation of Newton's law on the brane
and the study of graviton modes, which will be carried out in
Secs.~\ref{newton} and \ref{modes}, respectively.
 
For our purpose, we introduce a matter perturbation on the brane,
\begin{equation}
\label{perturb}
  \delta T_{00} = \delta(y) t_{00}(x),
\end{equation}
and solve the linearized gravity equations for this source.

The form \eqref{metric} of the background metric suggests to use the
time slicing formalism \cite{MTW} for calculating metric
perturbations, although we 
do not slice with respect to time, but with respect to the transverse
coordinate $y$.
Let us first give some useful formulae. In the time slicing formalism,
we split up the metric tensor as 
\begin{equation}
\label{gsplit}
  g_{ab} = \begin{pmatrix} g_{\mu\nu} & n_\nu \\
			   n_\mu & n_\mu n^\mu +n^2 \end{pmatrix},
  \qquad g^{ab} = \frac1{n^2} 
           \begin{pmatrix} n^2 g^{\mu\nu} + n^\mu n^\nu & -n^\nu \\
			   -n^\mu & 1 \end{pmatrix},
\end{equation}
where $a,b=0,1,2,3,5$, $x^5=y$, and $g_{\mu\nu}(x,y)$ are the induced
metrics in the hypersurfaces with internal coordinates $x^\mu$. 
The quantities $n$ and $n^\mu$
are called lapse function and shift vector, respectively, and are fixed
to their respective background values in the radiation gauge:
\begin{equation}
\label{fix}
  n^\mu =0, \qquad n=1.
\end{equation}
Thus, we consider a metric of the form
\begin{equation}
\label{metexpl}
  ds^2 = g_{\mu\nu} dx^\mu dx^\nu + dy^2.
\end{equation}
Then, the second fundamental form measuring the extrinsic curvature on
the hypersurfaces is given by
\begin{equation}
\label{H}
  H_{\mu\nu} =\frac1{2n} \left( \frac{\partial}{\partial y} g_{\mu\nu}
  - \nabla_\mu n_\nu - \nabla_\nu n_\mu \right) 
  = \frac12 \frac{\partial}{\partial y} g_{\mu\nu},
\end{equation}
where $\nabla_\mu$ is the covariant derivative on the hypersurfaces,
and the second equality holds in the radiation gauge. 

Einstein's equation is
\begin{equation}
\label{einstein}
  R_{ab} - \frac12 g_{ab} R = -g_{ab} \Lambda + 8\pi T_{ab},
\end{equation}
where $T_{ab} =\bar T_{ab} +\delta T_{ab}$, and $\bar T_{ab}$ is the
background from the brane, whose non-zero components are found from
eqn.\ \eqref{action} as
\begin{equation}
\label{tbar}
  \bar T_{\mu\nu} = \mp \frac{3k}{4\pi} \delta(y) g_{\mu\nu}.
\end{equation}
One can observe from eqns.\ \eqref{perturb} and \eqref{tbar} that
$T_{a5}=0$. Therefore, by virtue of the Gauss-Codazzi equations
\cite{MTW}, the normal and mixed components of eqn.\
\eqref{einstein} become
\begin{align}
\label{con1}
  \hat R + H^\mu_\nu H^\nu_\mu -H^2 &= 2 \Lambda,\\
\label{con2}
  \nabla_\mu H -\nabla_\nu H^\nu_\mu &= 0,\\
\intertext{respectively, where $H=H^\mu_\mu$, and $\hat R$ is the
intrinsic scalar curvature of the hypersurfaces. The tangential
components of eqn.\ \eqref{einstein} simply read}
\label{eqmot}
  R_{\mu\nu} -\frac12 g_{\mu\nu} R &= -g_{\mu\nu} \Lambda + 8\pi
  T_{\mu\nu}.
\end{align}
Eqn.\ \eqref{eqmot} is the equation of motion for $g_{\mu\nu}$,
whereas eqns.\ \eqref{con1} and \eqref{con2} are constraints.

We linearize eqns.\ \eqref{con1}--\eqref{eqmot} around the
background \eqref{metric}, for which purpose we use an induced
metric of the form 
\begin{equation}
\label{glin}
  g_{\mu\nu} = \e{\mp2k|y|} (\eta_{\mu\nu} + \gamma_{\mu\nu}),
\end{equation}
where $\gamma$ is a small perturbation compared to
$\eta$. The indices of $\gamma$ shall be raised and lowered 
using the Lorentz metric $\eta$. Some useful expressions for the
connections and curvatures are given in the appendix. 

Eqns.\ \eqref{con1}--\eqref{eqmot} take the forms 
\begin{align}
\label{con12}
  \e{\pm2k|y|}({\gamma^{\mu\nu}}_{,\mu\nu} -\Box \gamma)
  \pm 3k\sgn y \, \gamma_{,y} &=  0,\\
\label{con22}
  \frac12 \partial_y ( \gamma_{,\mu} - {\gamma^\nu}_{\mu,\nu} ) 
  &=0,\\
\notag
  \frac12 \left({\gamma^\mu}_{\rho,\mu\nu} +
  {\gamma^\mu}_{\nu,\mu\rho} - \Box \gamma_{\nu\rho} 
  -\gamma_{,\nu\rho} \right) -\frac12 \eta_{\nu\rho}
  \left({\gamma^{\mu\lambda}}_{,\mu\lambda} -\Box \gamma\right) &\\
\label{eqmot2}
  + \e{\mp2k|y|} \left[ -\frac12 \gamma_{\nu\rho,yy} + \frac12
  \eta_{\nu\rho} \gamma_{,yy} \pm 2k \sgn{y} (\gamma_{\nu\rho,y} -
  \eta_{\nu\rho} \gamma_{,y}) \right] &=  8\pi \delta T_{\nu\rho},
\end{align}
where the background has been cancelled using eqns.\ \eqref{lambda}
and \eqref{tbar}.

Let us start by solving the constraints. First, from eqn.\ \eqref{con22}
we find
\begin{equation}
\label{con2sol}
  {\gamma^\nu}_{\mu,\nu} = \gamma_{,\mu} +\xi_\mu(x),
\end{equation}
where $\xi_\mu$ are functions of the brane coordinates $x^\mu$ only. 
Secondly, after substituting eqn.\ \eqref{con2sol}, eqn.\ \eqref{con12}
leads to  
\begin{equation}
\label{gammaeq}
  \e{\pm2k|y|} {\xi^\mu}_{,\mu} \pm 3k \sgn{y} \gamma_{,y} =0.
\end{equation}
Thus, integrating eqn.\ \eqref{gammaeq} yields the trace $\gamma$ as 
\begin{equation}
\label{gamma}
  \gamma = -\frac1{6k^2} {\xi^\mu}_{,\mu} \left( \e{\pm2k|y|} -1
  \right),
\end{equation}
where we have used the residual gauge freedom to impose $\gamma=0$ on
the brane. We see from eqn.\ \eqref{gamma} that $\gamma$ is unbounded
for the RS background, if ${\xi^\mu}_{,\mu}\neq0$. Moreover,
this observation is independent of whether we use the residual gauge
freedom as indicated or not. We shall see later that we do not have
the choice of setting $\xi^\mu=0$, if a matter perturbation is present
on the brane. This indicates that, in the RS background, the
linear approximation is not consistent. However, this might be an
artifact of the particular choice of gauge. For further discussion of
this problem, see \cite{Garriga99,Giddings00}.

Finally, substituting eqns.\ \eqref{perturb}, \eqref{con2sol}
and \eqref{gamma} into the equation of motion \eqref{eqmot2},
we obtain the equation
\begin{equation}
\label{eqmot3}
\begin{aligned}
  \Box \gamma_{\nu\rho} +\partial_y \left( \e{\mp2k|y|}
  \gamma_{\nu\rho,y} \right) \mp 2k \sgn{y} \e{\mp2k|y|}
  \gamma_{\nu\rho,y} -\xi_{\nu,\rho} - \xi_{\rho,\nu} &\\
  + \frac13 \eta_{\nu\rho} {\xi^\mu}_{,\mu} +\frac1{6k^2} \left(
  \e{\pm2k|y|}-1\right) {\xi^\mu}_{,\mu\nu\rho} \pm \frac2{3k}
  \delta(y) \eta_{\nu\rho} {\xi^\mu}_{,\mu} &= -16\pi 
  \delta(y)t_{\nu\rho}.
\end{aligned}
\end{equation}
We can take the trace of eqn.\ \eqref{eqmot3} and find
\begin{equation}
\label{divxi}
  {\xi^\mu}_{,\mu} = \mp 8\pi k t,
\end{equation}
where $t=t^\mu_\mu$. 
Thus, as indicated earlier, the four-divergence ${\xi^\mu}_{,\mu}$ is
fixed by the content of matter perturbation on the brane. 

As the next step, we consider the discontinuity of eqn.\
\eqref{eqmot3} at $y=0$. One easily finds
\begin{equation}
\label{discont}
  -16 \pi t_{\nu\rho} = \gamma_{\nu\rho,y}|_{y=+0} 
  -\gamma_{\nu\rho,y}|_{y=-0} 
  \pm \frac{2}{3k} \eta_{\nu\rho} {\xi^\mu}_{,\mu}. 
\end{equation}
However, as the perturbation is symmetric around the brane, 
we need only look for solutions which are
even in $y$. Therefore, we can consider eqn.\ \eqref{eqmot3} in the
region $y>0$, and eqns.\ \eqref{discont} and \eqref{divxi} provide the
Neumann boundary condition
\begin{equation}
\label{bc}
  \gamma_{\nu\rho,y}|_{y=+0} = -8\pi \left(t_{\nu\rho} -\frac13
  \eta_{\nu\rho}t\right).
\end{equation}

Consider now eqn.\ \eqref{eqmot3} for $y>0$. First, let us choose the
vector $\xi^\mu$ as 
\begin{equation}
\label{xi}
  \xi_\mu = \mp 8\pi k \frac1{\Box} \partial_\mu t,
\end{equation}
which is consistent with the condition \eqref{divxi}. Then, we shall
write 
\begin{equation}
\label{ansatz} 
  \gamma_{\nu\rho} = \pm \frac{4\pi}{3k} \left[ 
  \frac1{\Box} t_{,\nu\rho} \left(\e{\pm2ky} -1 \right) 
  + 2 k^2 \left( \eta_{\nu\rho} \frac1{\Box} t - \frac4{\Box^2}
  t_{,\nu\rho} \right) \right] + \tilde \gamma_{\nu\rho} 
\end{equation}
in order to obtain from eqn.\ \eqref{eqmot3} the following homogeneous
equation for $\tilde\gamma_{\nu\rho}$:
\begin{equation}
\label{eqmot4}
  \Box \tilde\gamma_{\nu\rho} + \partial_y \left( \e{\mp2ky}
  \tilde\gamma_{\nu\rho,y} \right) \mp 2k \e{\mp2ky}
  \tilde\gamma_{\nu\rho,y} =0.
\end{equation}
Moreover, from eqn.\ \eqref{gamma} we find that the trace
$\tilde\gamma\equiv0$, and 
the Neumann boundary condition \eqref{bc} yields
\begin{equation}
\label{bc2}
  \tilde\gamma_{\nu\rho,y}|_{y=+0} = - 8\pi \left( t_{\nu\rho}
  -\frac13 \eta_{\nu\rho}t + \frac1{3\Box} t_{,\nu\rho} \right).
\end{equation}
Notice that a trivial (zero) solution to the homogeneous equation
\eqref{eqmot4} is not consistent with this boundary condition.

In order to solve eqn.\ \eqref{eqmot4}, let us Fourier transform with
respect to the brane coordinates and change variables to
$z=\e{\pm2ky}$. Then, eqn.\ \eqref{eqmot4} becomes 
\begin{equation}
\label{eqmot5}
  \left( z^2 \partial_z^2 -z \partial_z - \frac{p^2}{4k^2} z \right)
  \tilde\gamma_{\nu\rho}=0,
\end{equation}
whose solution can be expressed in terms of Bessel functions \cite{Gradshteyn}.

\section{Newton's Law} \label{newton}
In order to derive Newton's law on the brane, we have to look for a
unique solution to the linearized Einstein equations in the presence
of a static point source on the brane. In the last section, we
presented the general formalism of linearized gravity. Let us now
continue the solution for static point source.
In order to obtain the Newtonian limit, we have to calculate
$\gamma_{00}(x,0)$, since the gravitational potential is given by 
\begin{equation}
\label{potential}
  V= - \frac{m}2 \gamma_{00}.
\end{equation}

We need a second boundary condition for eqn.\ \eqref{eqmot4} in order
to obtain a unique solution. We shall simply use 
\begin{equation}
\label{bc3}
  \tilde\gamma_{\nu\rho}|_{y=+\infty} =0.
\end{equation}
For static potentials, we have $p_0=0$, and therefore
$p^2\ge0$ in eqn.\ \eqref{eqmot5}. In fact, we need only consider
$p^2>0$, as the solution for $p^2=0$ can be reconstructed as the limit
$p^2\to0$. The solution of eqn.\ \eqref{eqmot5} for $p^2>0$ is
\begin{equation}
\label{tgammasol}
  \tilde\gamma_{\nu\rho}(p,y) = c_{\nu\rho}(p) \e{\pm2ky} 
  \begin{cases} \K_2 \left(\e{\pm ky}|p|/k \right),\\
		\I_2 \left(\e{\pm ky}|p|/k \right), \end{cases}
\end{equation}
where the choice between the two possible solutions is dictated by
eqn.\ \eqref{bc3}. We easily see that this amounts to choosing the
solution with the $\K$ function for the RS background (upper
sign), and the solution with the $\I$ function for the alternative
background.  

Moreover, from eqn.\ \eqref{tgammasol} we find the first derivative,
\begin{equation}
\label{tgammay} 
  \tilde\gamma_{\nu\rho,y}(p,y) = \pm |p| c_{\nu\rho}(p) \e{\pm3ky} 
  \begin{cases}	-\K_1 \left(\e{\pm ky}|p|/k \right),\\
		\I_1 \left(\e{\pm ky}|p|/k \right), \end{cases}
\end{equation}
which, combined with the boundary condition \eqref{bc2}, yields the
coefficients 
\begin{equation}
\label{c}
  c_{\nu\rho} = \frac{8\pi}{|p|} \left[ t_{\nu\rho}
  -\frac13 t \left( \eta_{\nu\rho}-\frac{p_\nu p_\rho}{p^2} \right)
  \right]  
  \begin{cases} 
  [\K_1(|p|/k)]^{-1} \quad 
  &\text{for $\sigma<0$,}\\
  {}[\I_1(|p|/k)]^{-1} &\text{for $\sigma>0$.} 
  \end{cases}
\end{equation}

Thus, inserting eqns.\ \eqref{tgammasol} and \eqref{c} into eqn.\
\eqref{ansatz}, we obtain the solution for the metric perturbation
\begin{equation}
\label{gammap}
\begin{aligned}
  \gamma_{\nu\rho}(p,y) &= \pm \frac{4\pi}{3k} \left[ 
  \frac{p_\nu p_\rho}{p^2} \left(\e{\pm2ky} -1 \right) 
  - 2 k^2 \left( \frac{\eta_{\nu\rho}}{p^2} - \frac{4p_\nu p_\rho}{p^4}
  \right) \right] t \\
  &\quad + \frac{8\pi}{|p|} \left[ t_{\nu\rho}
  -\frac13 t \left( \eta_{\nu\rho}-\frac{p_\nu p_\rho}{p^2} \right)
  \right] 
  \begin{cases} \e{2ky}
  \frac{\K_2\left(\e{ky}|p|/k\right)}{\K_1(|p|/k)} \quad 
  &\text{for $\sigma<0$,}\\
  \e{-2ky}
  \frac{\I_2\left(\e{-ky}|p|/k\right)}{\I_1(|p|/k)}
  &\text{for $\sigma>0$.} \end{cases}
\end{aligned}
\end{equation}

Let us now use a static point source, $t_{00}(p)= 2\pi\delta(p_0) a$,
where $a=M/\mpl{5}^3$, and solve for $\gamma_{00}(x,0)$. Consider first the
RS background. We can use the recursion formula for modified
Bessel functions,
 \[ \K_2(z) =\frac2z \K_1(z) + \K_0(z), \]
in order to separate the divergent term for $|p|\to0$ in the Fourier
integral. Then, from eqn.\ \eqref{gammap} we find
\begin{align}
\notag
  \gamma_{00}(x,0) &= - \frac{2ka}{3r}
  + \frac{8ka}{3r}
  + \lim_{y\to0} \int \frac{d^3 p}{(2\pi)^3} \e{-ip\cdot x}
  \frac{16\pi a}{3|p|}
  \frac{\K_0(\e{ky}|p|/k)}{\K_1(|p|/k)}\\
\label{gammax1}
  &= \frac{2ka}{r} + 
  \frac{8a}{3\pi r^2}
  \lim_{y\to0} \int\limits_0^\infty ds\, \sin s\, 
  \frac{\K_0(\e{ky}s/kr)}{\K_1(s/kr)}.
\end{align}
It is interesting to note that the inhomogeneous terms of the solution
would yield Newton's law with a wrong sign, but the homogeneous part,
whose presence is necessary because of the Neuman boundary condition,
takes care of this. In fact, the importance of the boundary conditions
for obtaining the gravity on the brane has already been pointed out in
\cite{Dick00}. Moreover, we observe that the integral in eqn.\
\eqref{gammax1} is well-defined only for $y>0$. It is for this reason that
we take the limit $y\to0$ after the integration. The non-zero $y$ acts
as a regulator of the integral for large $s$. The second
term in eqn.\ \eqref{gammax1} is a correction to Newton's law, which
we shall demonstrate in a moment. 

Let us consider now the alternative background. From eqn.\
\eqref{gammap} we obtain 
\begin{align}
\notag
  \gamma_{00}(x,0) &= \frac{2ka}{3r} + 
  \lim_{y\to0} \int \frac{d^3 p}{(2\pi)^3} \e{-ip\cdot x}
  \frac{16\pi a}{3|p|}
  \frac{\I_2(\e{-ky}|p|/k)}{\I_1(|p|/k)}\\
\label{gammax2}
  &= \frac{2ka}{3r} + 
  \frac{8a}{3\pi r^2}
  \lim_{y\to0} \int\limits_0^\infty ds\, \sin s\, 
  \frac{\I_2(\e{-ky}s/kr)}{\I_1(s/kr)}.
\end{align}
The first term in eqn.\ \eqref{gammax2} represents Newton's law, and
the second term corrections, as we shall demonstrate now. 

Consider an integral of the form $\int_0^\infty ds\, \sin s\, f(s/z,y)$,
where $f$ is a differentiable and integrable function. From
\cite{Gradshteyn} we know that the
integrands in eqns.\ \eqref{gammax1} and \eqref{gammax2} satisfy this
property for any $y>0$. Given the integrability of $f$, 
we can rewrite the integral as 
\begin{equation}
  \int\limits_0^\infty ds\, \sin s\, f(s/z,y) =
  \sum\limits_{k=0}^\infty \int\limits_{-\pi}^\pi ds\, (-\sin s)\,
  f\left(\frac{\pi(2k+1)+s}z,y\right).
\end{equation}
At this point, we can take the limit $y\to0$, and we shall write
$f(x,0)=f(x)$. For large $z$, the argument of $f$ will change little in one
period of the $\sin$ function, and we can write
\begin{align*}
  \int\limits_0^\infty ds\, \sin s\, f(s/z) &\approx 
  -\sum\limits_{k=0}^\infty \int\limits_{-\pi}^\pi ds\, \sin s\, 
  \left[ f\left(\frac{\pi(2k+1)}z\right) + \frac{s}z
  f'\left(\frac{\pi(2k+1)}z\right) \right]\\   
  & \approx - \sum\limits_{k=0}^\infty
  \frac{2\pi}z f'\left(\frac{\pi(2k+1)}z\right).
\end{align*}
Here, we observe that the $z\to\infty$ limit exists, namely it is
just the integral
\[ -\int\limits_0^\infty dx\, f'(x) = f(0) -f(\infty).\]
For both integrands under consideration we have
$f(0)=0$ and $f(\infty)=1$ (for $y>0$ we would have the stronger
$f(\infty)=0$). Thus, we find that the second terms in eqns.\
\eqref{gammax1} and \eqref{gammax2} go to zero at least as fast as
$1/r^2$ for large $r$ and are in fact corrections to Newton's law. 

Finally, from eqn.\ \eqref{potential} we obtain the gravitational
potential 
\begin{equation}
\label{V}
   V(r) = -\frac{kmM}{\mpl{5}^3} \begin{cases} 1/r
   +\mathcal{O}(1/r^2) \quad &\text{for $\sigma<0$,}\\
   1/(3r) +\mathcal{O}(1/r^2) &\text{for $\sigma>0$.} \end{cases}
\end{equation}
Thus, one can read off the four-dimensional Planck masses as
\begin{equation}
\label{mpl}
  \mpl{4}^2 = \begin{cases} \mpl{5}^3/k 
  \quad &\text{for $\sigma<0$,}\\
  3\mpl{5}^3/k  &\text{for $\sigma>0$.} 
\end{cases}
\end{equation}
Our result for the RS background is in agreement with the result given
in \cite{Randall99a}.

\section{Graviton Modes} \label{modes}
It is our objective in this section to study graviton modes. The
equation of motion for gravitons in the gauge $\delta g_{a5}=0$ is the
homogeneous equation \eqref{eqmot4} with the boundary condition
\begin{equation}
\label{bcmodes}
  \gamma_{\nu\rho,y}|_{y=0} =0.
\end{equation}
This boundary condition stems from eqn.\ \eqref{bc} and implies that
we consider only modes which are even in $y$. This would be natural
for the orbifold $S^1/\mathbb{Z}_2$, but in general one
would have to match a solution for $y>0$ with a solution for $y<0$. 

Let us now give the graviton solutions. As we consider only modes
which are even in $y$, we restrict ourselves to $y\ge0$. Using eqn.\
\eqref{eqmot5}, the solutions to eqn.\ \eqref{eqmot4} satisfying the
boundary condition \eqref{bcmodes} are 
\begin{equation}
\label{modessol}
  \gamma_{\mu\nu}(p,y) = c_{\mu\nu}(p) \begin{cases}
  \e{\pm2ky} \left[\I_2(\alpha\e{\pm ky}) \K_1(\alpha) +
  \K_2(\alpha\e{\pm ky}) \I_1(\alpha)\right] \quad&\text{for
  $p^2>0$,}\\
  1 &\text{for $p^2=0$,}\\
  \e{\pm2ky} \left[\J_2(\alpha\e{\pm ky}) \N_1(\alpha) -
  \N_2(\alpha\e{\pm ky}) \J_1(\alpha)\right] \quad&\text{for $p^2<0$.}
  \end{cases}
\end{equation}
Here, we have set $\alpha=\sqrt{|p^2|}/k$.

Let us discuss the normalizability of these modes. For the RS
background, one finds that space-like modes ($p^2>0$) are not normalizable,
since $\I_2$ diverges as $\e{\e{y}}$ for large $y$. On the other hand,
zero-like and time-like modes are normalizable. 

On the other hand, for the alternative background the $y$ integral 
diverges in all three cases just as the volume
integral. Thus, just as in flat space, we might assume that we can
form wave packets which describe normalizable wave functions. 
Alternatively, one might regularize integrals like the norm of
wavefunctions or the effective action by dividing by the total volume
of space. We would like to leave these points open to further research.

\section{Conclusions} \label{conc}
In this paper, we have studied the geodesics and derived Newton's law
for the Randall-Sundrum and an alternative brane background. We found
that matter will be expelled from the brane in the RS
background. A similar behaviour was observed by Rubakov \emph{et al.}\
\cite{Rubakovprivate} in a
different context. Therefore, the RS background is classically unstable, and
it must be supplemented with a mechanism for confinement of matter. On
the other hand, gravity provides a natural mechanism for the trapping of
matter in the alternative background. It would be interesting to study
a realization of the alternative background in supergravity. The RS
background in supergravity has been discussed in
\cite{Behrndt99,Youm99,Youm00a,Youm00b,Kallosh00}. 

Our derivation of gravity on the brane revealed the validity of Newton's law
to leading order in both backgrounds, but the corresponding Planck
masses on the brane are different. Moreover, we found exact formulas,
eqns.\ \eqref{gammax1} and \eqref{gammax2}, for the corrections. 

In the RS background there are no normalizable space-like modes, which
certainly is in favour of this background. For the alternative background, all modes
are non-integrable, but a thorough discussion of the confinement or
non-confinement of gravity remains an open problem. 
It seems that the confinement of gravity cannot be discussed separately
from the confinement of matter.

\section*{Acknowledgements}
This research was partly supported by NSERC.
I. V. is grateful to the Physics Department of Simon Fraser University
for its kind hospitality.
Moreover, we are grateful to I. Ya.\ Aref'eva and M. G. Ivanov for their
help in the early stage of this work and to A. Linde and V. A. Rubakov
for their critical remarks.  

\begin{appendix}
\section*{Appendix}
We state here some expressions for the connections and curvature up to
first order in the perturbations $\gamma_{\mu\nu}$.  
The only non-zero connections for the metric $g_{ab}$ are 
\label{useful}
\begin{align}
  {\Gamma^\mu}_{\nu\lambda} &= \frac12 \left( 
  {\gamma^\mu}_{\nu,\lambda} + {\gamma^\mu}_{\lambda,\nu}  
  - {\gamma_{\nu\lambda}}^{,\mu} \right),\\
  {\Gamma^y}_{\nu\lambda} &=\pm k \sgn y \, g_{\nu\lambda} - \frac12
  \e{\mp2k|y|} \gamma_{\nu\lambda,y},\\
  {\Gamma^\nu}_{\lambda y} = {\Gamma^\nu}_{y\lambda} &= \mp k\sgn{y}
  \delta^\nu_\lambda +\frac12 {\gamma^\nu}_{\lambda,y}.
\end{align}
Moreover, we find from eqn.\ \eqref{H}
\begin{equation}
  H^\mu_\nu = \mp k\sgn{y} \delta^\mu_\nu + \frac12
  {\gamma^\mu}_{\nu,y},
\end{equation}
and some expressions for the curvatures are
($\Box=\eta^{mu\nu}\partial_\mu\partial_\nu$)
\begin{align}
  \hat R_{\nu\rho} &= \frac12 \left( {\gamma^\mu}_{\nu,\rho\mu} +
  {\gamma^\mu}_{\rho,\nu\mu} - \gamma_{,\nu\rho} -\Box
  \gamma_{\nu\rho} \right),\\
  \hat R &= \e{\pm 2k|y|} \left( {\gamma^{\mu\nu}}_{,\mu\nu} -\Box
  \gamma \right),\\ 
  {R^y}_{\nu y\rho} &= -k^2 g_{\nu\rho} \pm 2k \delta(y) g_{\nu\rho} +
  \e{\mp 2k|y|} \left(\pm k \sgn{y} \gamma_{\nu\rho,y} -\frac12
  \gamma_{\nu\rho,yy} \right),\\
  R_{\mu\rho} &= \hat R_{\mu\rho} + H^\nu_\mu H_{\nu\rho} - H
  H_{\mu\rho} + {R^y}_{\mu y\rho},\\ 
  R &= \hat R -20 k^2 \pm 16 k \delta(y) \pm 5k \sgn{y} \gamma_{,y}
  -\gamma_{,yy}.
\end{align}
\end{appendix}

%\bibliographystyle{art}
%\bibliography{strings,misc,rs}

\end{document}